\documentclass[12pt,epsf]{article}
\input epsf.tex
\def\be{\begin{equation}} \def\ee{\end{equation}}
\def\bi{\begin{itemize}} \def\ei{\end{itemize}}
\def\bea{\begin{eqnarray}} \def\eea{\end{eqnarray}} \def\ba{\begin{array}}
\def\ea{\end{array}} \def\ben{\begin{enumerate}} \def\een{\end{enumerate}}
  \def\nnu{\nonumber}

\newcommand{\eqn}[1]{(\ref{#1})}

\newcommand{\prl}[3]{Phys. Rev. Lett. {\bf#1} ({#2}) {#3}}

\newcommand{\prd}[3]{Phys. Rev. {\bf D#1} ({#2}) {#3}}
\newcommand{\hepth}[1]{{\tt [arXiv:{#1}[hep-th]]}}

\def\br{\nonumber\\}

\begin{document}
{}~
\hfill\vbox{\hbox{hep-th/1305.3784} 
\hbox{\today}}\break

\vskip 3.5cm
\centerline{\large \bf
Lifshitz to AdS flow with interpolating $p$-brane solutions} 

\vskip .5cm

\vspace*{.5cm}

\centerline{\bf  Harvendra Singh}

\vspace*{.5cm}
\centerline{ \it  Theory Division} 
\centerline{ \it  Saha Institute of Nuclear Physics} 
\centerline{ \it  1/AF Bidhannagar, Kolkata 700064, India}
\vspace*{.25cm}

\vspace*{.5cm}

\vskip.5cm

\vskip1cm

\centerline{\bf Abstract} \bigskip

In continuation with our studies of Lifshitz like D$p$-brane solutions, 
we propose a class of $1/4$ BPS
supersymmetric interpolating solutions which interpolate between IR
Lifshitz solutions and UV AdS solutions smoothly. We demonstrate 
properties of these classical solutions near the two fixed points. 
These interpolating solutions are then used to calculate the
entanglement entropies of strip-like subsystems. 
With these bulk solutions the entropy functional also gets modified.
We also make a curious observation about the electric-magnetic 
duality and the thermal entropy of the Hodge-dual Lifshitz D$p$ brane
systems. 

\vfill 
\eject

\baselineskip=16.2pt


\section{Introduction}

Recently a significant amount of work is being carried out \cite{son}-
\cite{narayan13},
 on the construction of the string duals of some strongly coupled quantum
systems near the critical fixed points, exhibiting 
Lifshitz type scaling symmetries \cite{kachru}
\be 
t\to \lambda^a t, ~~~~ x^i\to \lambda x^i \ .
\ee
Namely the time and space coordinates in the CFT do scale {\it asymmetrically}.
Some of these systems exhibit a non-fermi liquid or strange metallic behaviour 
at ultra low temperatures, see for details \cite{takaya11,subir11}. 
There are also issues related to 
the entanglement entropy of the  quantum subsystems \cite{RT, takaya11}. 
The  entanglement
entropy of the subsystems can  be defined geometrically 
as the area of a minimal surface within the bulk, with specific boundary 
conditions \cite{RT}.  Recently,  a class of
 Lifshitz and Schro\'dinger type spacetimes have  been  
 constructed
in type II string theory and M theory, 
exhibiting a fixed amount of supersymmetry \cite{hs10,hs10a}. 
The Lifshitz like solutions  have also been shown 
to arise in 
\cite{naray, kim, Narayan:2011az,Chemissany:2011mb}  
and from intersecting D-branes in \cite{dey2012}. 
Our main
focus in this work is a class of $1/4$ BPS Lifshitz D$p$ solutions 
\cite{hs10,hs10a}, which can be generically  obtained
as vanishing horizon  double limits of the boosted 
black $p$-branes vacua \cite{hs10a}.   
The supersymmetry  makes these Lifshitz 
solutions more interesting, because
we can find more definitive predictions about the boundary nonrelativistic
CFT. Let us note that some of
these Lifshitz IR solutions have problems in the  
UV region. 
In the paper \cite{hsuv},  an specific  resolution of
the UV problem was attempted for D3-brane Lifshitz solutions. 
Particularly, the solutions were modified 
such that they remain well behaved classical geometries even in the UV region.
Here in this article we extend that particular approach to all  Lifshitz D$p$ 
solutions given in \cite{hs10a}. 
We write down a new class of $1/4$ 
supersymmetric solutions which can interpolate between (IR)
Lifshitz solutions and (UV) AdS solutions smoothly. We demonstrate 
various properties of these classical solutions in 
the IR and UV asymptotic regions. 
These interpolating bulk solutions are then used to calculate the
entanglement entropy of strip-like subsystems of the boundary CFT. 
In general, the entropy functional gets modified.
We also make a curious observation about the effect of electric-magnetic 
duality on the thermal entropy of the electric/magnetic (Hodge) dual
Lifshitz solutions. 
For example, 
the entropy of the near extremal Lifshitz D$p$-brane goes as
\be 
S_{(p)}\sim T^{1\over \tilde p}
\ee
where $\tilde p$ is nothing but the number of spatial world-volume 
directions of the corresponding (magnetic) dual Lifshitz D$\tilde{p}$-brane.
The same relation holds good for the pair of non-extremal 
Lifshitz  M2 and M5-branes. 

The paper is planned in the following way. In the section-2 we review the
basic properties of the maximally supersymmetric $AdS\times S$ vacua in 
type II string theory. In section-3 we study the $1/4$ BPS Lifshitz 
D$p$-brane vacua and obtain expressions for
 their thermal quantities at finite temperature. We do show  how  various
thermodynamical quantities behave when vanishing temperature limit is taken. 
We explore the 
effect of electric-magnetic duality on the thermal entropy. 
In  section-4 we write down new interpolating solutions which are well behaved in the UV. 
We obtain the
entanglement entropy using these smooth interpolating  solutions. 
The entanglement
entropy expression matches with the
 recent works \cite{narayan13,narayan12} for the strip like subsystems.
 The conclusions are given 
in the section-5.

\section{D$p$-branes and relativistic CFTs}
The maximally supersymmetric near horizon D$p$-brane  solutions  are 
given by \cite{itzhaki}
\bea\label{sol2de1}
&&ds^2_{AdS}=R_p^2r^{p-3\over2}\bigg[ 
r^{5-p}[  (dx^{-})^2 -dx^{+}dx^{-}
+d\vec x_{(p-1)}^2]  
+{dr^2\over  r^2}  + d\Omega_{(8-p)}^2 \bigg]\ , \br
&& e^\phi=(2\pi)^{2-p}g_{YM}^2 
R_p^{3-p} r^{(7-p)(p-3)\over4} 
\eea 
along with a suitable $(p+2)$-form field strength
\be
F_{p+2}=(7-p)R_p^{2p-2}r^{6-p} 
dr\wedge dx^+\wedge dx^-\wedge [dx_{(p-1)}]\ee
for the electric type D$p$-branes $(p<3)$ and a $(8-p)$ form
\be F_{8-p}=(7-p) R_p^4\, \omega_{8-p}\ee
for the magnetic type (hodge dual) D$p$-branes $(p>3)$.
Specially for D3-brane case
we  have $F_5=4 (1+\star)\omega_5$, which is self-dual 5-form field strength. 
We have introduced $x^+,x^-$ as  lightcone coordinates along the 
world volume of the branes,  
and $\vec x_{(p-1)}$ represents other $(p-1)$ spatial directions parallel 
to the D$p$-brane, and as usual $r$ is the radial (holographic) coordinate.
The interpretation of various parameters can be found in 
\cite{itzhaki} and also given in  \cite{hs10a}.

\noindent{\it A Proposal}:

One should note that, in these conformally  $AdS_{p+2}\times S^{8-p}$ 
solutions, we have  taken a slightly modified  $AdS$ metric elements: 
$r^{5-p}[  (dx^{-})^2 -dx^{+}dx^{-}
+d\vec x_{(p-1)}^2]  
+{dr^2\over  r^2}$. Namely we have introduced a constant  
$g_{--}$ component. Doing this is actually harmless as it 
still remains an AdS geometry. The constant $g_{--}$ term
can be reabsorved by a coordinate shifts like
$x^+\to x^+ + x^-$, if  the need arises. However, certain 
global symmetries of the metric,
such as the  lightcone boost $x^-\to \lambda x^-,~
x^+\to {1\over \lambda} x^+$, are spontaneously broken in this 
new modified frame. The inclusion of   $g_{--}$ 
component in these solutions is
 useful in the following way. 
We shall be considering  (nonrelativistic)
Lifshitz-like solutions, having nontrivial $g_{--}$ deformations. 
In these solutions,
 we shall take   $x^-$ to be mostly a compact direction. 
Note that, when $x^-$ is compactified, 
 in order to trust our classical string metric,
it makes sense to keep $g_{--}$  finite 
instead of taking it to be vanishing, also see comments in Ref.\cite{malda}. 
Of course, in the `shifted' $(x^+,x^-)$ frame, as in \eqn{sol2de1}, 
the DLCQ of boundary 
CFT will have a slightly changed energy-mass relationship; 
see the discussion  in the appendix of \cite{hs10a}.
For the case of D3-branes, such UV geometry was specifically proposed as 
a resolution of `UV problem' of the $a=3$  Lifshitz solutions of \cite{hs10a}. 
We must emphasize that all our Lifshitz 
solutions, as given below in section-3, are good only in some intermediate range of $z$, 
while they do have  problem 
in extreme UV region, i.e. as we get near to the AdS boundary.  
We  propose  that,
all those Lifshitz solutions $(p<5)$ can  be modified so as to 
include the spacetime in eq.\eqn{sol2de1} as the asymptotic metric in  UV.

Let us  redefine the radial coordinate 
\be 
r^{p-5}=z^2 ~~~~~~~~~~{\rm for}~ p\ne 5
  \ee
Similar, redefinition of the radial coordinate
 can be done for D5-brane 
separately, if required.
With $z$ as holographic coordinate and some 
 scaling of the brane coordinates the above solutions 
can be brought to the form 
\bea\label{sol2d03}
&& ds^2=R_p^2 z^{p-3\over p-5} \bigg[ 
\{ {(dx^{-})^2\over z^2} +{-dx^{+}dx^{-}
+d\vec x_{(p-1)}^2\over z^2}  
+{4\over (5-p)^2} {dz^2\over  z^2}\}  + d\Omega_{(8-p)}^2 \bigg] \br
&& e^\phi=(2\pi)^{2-p}g_{YM}^2R_p^{3-p}{ z^{(7-p)(p-3)\over 2( p-5)}}
\eea 
along with the $(p+2)$-form flux.
One can find that under the  dilatations the 
 coordinates would rescale as
\be\label{d1}
z\to \xi z , ~~~~
x^{\pm}\to \xi x^{\pm}, ~~~\vec x\to \xi \vec x
\ee
while  the dilaton and 
the string metric  in \eqn{sol2d03} conformally rescale as
\bea\label{d2}
g_{MN}\to \xi^{p-3\over p-5} g_{MN},
~~~e^\phi\to \xi^{(7-p)(p-3)\over2(p-5)}e^\phi
\eea
Note this latter conformal rescaling is  
the standard  Weyl rescaling behaviour, 
 of non-conformal D$p$-branes AdS solutions \cite{itzhaki}, 
giving rise to the RG flow in the boundary CFT. 
From  Eq.\eqn{d1} the  dynamical exponent of time 
is $a\equiv a_{rel}=1$, 
so that the boundary theories
 are  $(p+1)$-dimensional
`relativistic' CFT$_{(p+1)}$ with sixteen supercharges. 
Note, once $x^-$ is taken to be a coordinate on a circle, the 
boundary CFT becomes a DLCQ theory and is a $p$-dimensional theory. 
While the compactification of the bulk solution 
\eqn{sol2d03} along $x^-$ and $S^{8-p}$, 
 results in $(p+1)$-dimensional 
(Einstein) metric given as
\bea\label{sol2d}
ds^2_{p+1}
&\sim& z^{({p-5\over p-1}+{p-3\over p-5})} \bigg[ 
 -{(dx^{+})^2 \over z^2}+{d\vec x_{(p-1)}^2\over z^2}  
+{4\over (5-p)^2} {dz^2\over  z^2}   \bigg] \br 
&= &z^{2(p^2-7p+14)\over (p-1) (p-5)} \bigg[ 
 -{(dx^{+})^2 \over z^2}+{d\vec x_{(p-1)}^2\over z^2}  
+{4\over (5-p)^2} {dz^2\over  z^2}   \bigg] 
\equiv z^{2\theta \over d} 
 ds^2_{AdS_{p+1}}\ .  
\eea 
From where we can read the hyperscaling parameter, 
as it is known now, to be
\be\label{thea1}
 \theta={p^2-7p+14 \over p-5}\equiv \theta_{rel} .
\ee
Note that, $d\equiv p-1$  gives the total number of spatial directions
of the boundary CFT$_{p}$. 
Let us  mention here that there is also  
a running $(p+1)$-dimensional dilaton field
\be
 e^{-2\phi_{(p+1)}}\sim z^{p-5\over2}
\ee
 as well as other 
form fields arising out of reduction of $(p+2)$-form field strength. 
These  solutions are  extremal solutions. 

\subsection{ The thermal entropy of the relativistic theory}
In order to know the thermal behaviour of the boundary CFT, one 
incudes black holes in the bulk anti-de Sitter geometry. In our coordinates
the near extremal D$p$ solutions are
\bea\label{int031a}
ds^2 &=&R_p^2 z^{p-3\over p-5} \bigg[ 
\{-{(f-1)(dx^{+})^2\over4 z^2 }+
{-dx^- dx^++(dx^-)^2+d\vec x_{(p-1)}^2\over  z^2}  
+{4\over (5-p)^2} {dz^2\over f z^2}\}
  + d\Omega_{(8-p)}^2 \bigg]\nonumber \\
&=&R_p^2 z^{p-3\over p-5} \bigg[ 
\{-{f(dx^{+})^2\over4 z^2 }+ 
{d\vec x_{(p-1)}^2\over  z^2}  
+{4\over (5-p)^2} {dz^2\over f z^2}\}+
{1\over z^2}(dx^- -{1\over 2}dx^{+})^2
  + d\Omega_{(8-p)}^2 \bigg]\nonumber \\
\eea
where function $$f=1-\left({z\over z_{0}}\right)^{2p-14\over p-5}$$ 
vanishes at $z=z_0~(z_0>0)$ as it is the location of the horizon. 
As usual with black hole  D$p$-branes,
the dilaton and other flux form fields remain unchanged.
Corresponding  thermal CFTs have a definite temperature
behaviour. For example,
the entropy density, $s$, of the relativistic theories 
 \cite{itzhaki}
\be
s\equiv{S\over V_d}
\sim  2\pi r^- T^{p-9\over p-5}, ~~~~~ T\sim z_0^{-1}
\ee
where $V_d$ is the volume of the $d$-dimensional spatial ensemble box.
There is also a
 chemical potential $\mu \sim {1\over 2 r_-}$, which is trivial, as the
corresponding charge density is vanishing. This is simply an artefact of our
coordinate choice (shifted lightcone frame). This could be undone by 
a gauge choice, but we do not worry about it here.
 Thus the system  is still a 
canonical ensemble with a fixed number of particles. 
Using the expression for $\theta_{rel}$ given above, 
 entropy is also  expressible as
\be
s\sim T^{p-1-\theta_{rel}}
\equiv T^{d-\theta_{rel} \over a_{rel}}
\ee
Note that 
the dynamical exponent of time coordinate in relativistic solutions is simply
unity. 
Thus literally  there is 
a hyperscaling violation as $(\theta_{rel}\ne 0)$
 in these relativistic systems too,
 due to the nontrivial conformal factors in the  metrics. 
This is an all familiar terrain so far.
We prepare a table of the corresponding  CFT data in the table 
\eqn{Table1}.
\begin{table}[h]
\begin{center} 
\begin{tabular}{ccccc}
\hline D$p$-brane &$d$ & 
$a_{rel}$& $ \theta_{rel}$ & $s\sim T^\alpha $ \\ \hline 
 \hline
1 &0& 1 & $ -2$ &$T^2$ \\ \hline
2 &1& 1 & $-{4\over 3}$ &$T^{7\over 3}$ \\ \hline
3 &2& 1 &$ -1$ &$T^3$ \\ \hline
4 &3& 1 & $-2$ &$T^5$ \\ \hline
\end{tabular} 
\caption{\label{Table1} Dynamical scaling exponents and $\theta$ parameter
arising out  of the relativistic D$p$ brane solutions}
\end{center}
\end{table}
The 
 exponent $\alpha$ of the $T$ in the entropy expression
 increases with the increase 
in the dimensionality of the relativistic ensemble.

 \section{ ${1\over 4}$-BPS Lifshitz  D$p$-branes }

The Lifshitz like D$p$ solutions with eight supersymmetries
are given by 
\cite{hs10,hs10a}
\bea\label{sol3}
&&ds^2_{lif}=R_p^2 z^{p-3\over p-5} \bigg[ 
\{ {\beta^2\over z^{4/(p-5)}} (dx^{-})^2 +{-dx^{+}dx^{-}
+d\vec x_{(p-1)}^2\over z^2}  
+{4\over (5-p)^2} {dz^2\over  z^2} \}  + d\Omega_{(8-p)}^2 \bigg] ,\br
&& e^\phi=(2\pi)^{2-p}(g_{YM})^2 R_p^{3-p}{ z^{(7-p)(p-3)\over 2(p-5)}}
\eea 
with the $(p+2)$-form flux, given above (for $p\ne 5$). 
Here $\beta$ is arbitrary scale parameter
and can be absorbed by  scaling the lightcone coordinates.
These solutions can simply be obtained by employing `vanishing horizon 
double limits' of the boosted black D$p$-branes solutions \cite{hs10,hs10a}. 
These could also be described as conformally AdS spacetimes with plane wave, 
having momentum along $x^-$.  
In these Lifshitz like solutions  the light cone coordinates
do scale  {\it asymmetrically} under the dilatations 
\be\label{sol4}
z\to \xi z , ~~~~
x^{-}\to \xi^{2-a}  x^{-},~~~~
x^{+}\to \xi^{a} x^{+},~~~\vec x\to \xi \vec x
\ee 
with the dynamical exponent of time 
 $a=a_{lif}={2p-12\over p-5}$.  At the same time the dilaton field and 
the metric  in eq.\eqn{sol3} conformally rescale as in eq.
\eqn{d2}.
These Lifshitz solutions  \eqn{sol3}, 
on explicit compactifications along $x^{-}$ and $S^{8-p}$, 
generically
give rise to $(p+1)$-dimensional noncompact Lifshitz metrics 
(in Einstein frame) 
\be\label{pok9a}
ds^2_{lif_{p+1}}\sim
z^{2(p^2-6p+7)\over (p-1)(p-5)}\left( -{(dx^{+})^2\over \beta^2 
z^{2 a_{lif}}}
+{d\vec{x}_{p-1}^2  +dz^2\over  z^2}\right) ,
\ee
with 
\be\label{sol5}
  a_{lif}={2p-12\over p-5}\ee
Thus the hyperscaling parameter 
\be\label{sol6}
\theta_{lif}={p^2-6p+7\over p-5},
\ee
for all $0<p\le 6$ but $p\ne 5$. 
Note that $\theta$ is never vanishing 
in these Lifshitz solutions \eqn{pok9a} or
in the relativistic  solutions \eqn{sol2d}. In fact, we generally find that
\be
\theta_{lif}>
\theta_{rel}\ .
\ee
for all $p$ cases. The physically interesting cases are with $p=2,3,4$,
and  they all satisfy 
$a\ge {\theta\over d}+1$. The corresponding boundary 
nonrelativistic CFTs do have spatial dimensions $d=1,2,3$ respectively. 
These systems could hopefully be realized in nature.

\subsection{ The thermal  entropy of a Lifshitz system}
The thermal behavior of the entropy at the Lifshitz 
fixed points could be studied if we  consider black holes
in the Lifshitz  solutions \eqn{sol3}. It
is described by the following type of black hole solutions \cite{hs10a} 
\bea\label{sol7}
ds^2_{Lif}&=&R_p^2 z^{p-3\over p-5} \bigg[ 
\{-{f(dx^{+})^2\over4 z^2 g}+ 
{d\vec x_{(p-1)}^2\over  z^2}  
+{4\over (5-p)^2} {dz^2\over f z^2}\}\br
&&~~~~~~~~+{g\over z^2}(dx^- -{1+f\over 4g}dx^{+})^2
  + d\Omega_{(8-p)}^2 \bigg]\nonumber \\
\eea
where functions $$f=1-\left({z\over z_{0}}\right)^{2p-14\over p-5}$$ 
while $ g(z)\equiv 
{1\over4}({z\over z_{IR}})^{2(p-7)\over p-5}$, where $z_{IR}>0$
is some an intermediate IR scale.
Also $z_0> z_{IR}$ is the black hole horizon. 
Note that, 
the dilaton and the $(p+2)$-form field strengths remain same as in 
the relativistic solutions. 
 
The thermal entropy of the system (not the entanglement entropy) 
is obtained by estimating the area of the black hole horizon, It
can be summarised by the same type of expression 
as in the relativistic case, namely
\be\label{sol8}
s\sim (2\pi r^-) T^{d-\theta_{lif}\over a_{lif}}, ~~~~T\sim z_0^{- a_{lif}}\ .
\ee
\begin{table}[h]
\begin{center} 
\begin{tabular}{ccccc}
\hline D$p$-brane&$d$ & $a_{lif}$& $ \theta_{lif}$ & $s $ \\ \hline 
 \hline
D1&0 & ${5\over 2}$ &  ${-1 \over 2}$ &$T^{1\over5}$ \\ \hline
D2&1 & ${8 \over 3}$ & ${1\over3}$ &$T^{1\over4}$ \\ \hline
D3&2 & 3 & 1 &$T^{1\over3}$ \\ \hline
D4&3 & 4 & 1 &$T^{1\over2}$ \\ \hline
D5&4 & 1 & 3 &$T^{1}$ \\ \hline
\end{tabular} 
\caption{\label{Table2} Dynamical scaling exponents of the 
Lifshitz solutions}
\end{center}
\end{table}
While the chemical potential and the charge density is given by
\be\label{pol3}
\mu_{_N}\sim {1\over r_{-}}\left({z_{IR}\over z_{0}}\right)^{2p-14\over p-5},
~~~~\rho\sim r_{-}^2  {z_{IR}}^{2p-14\over 5-p}
\ee
where $d\equiv p-1$ is the number of spatial dimensions of the CFT.
Note, the thermal behaviour of the system, particularly
in very low temperature limit $T\to 0$ (as $r_h\equiv 1/z_0\to 0 $) 
can be determined 
at a fixed charge density ($z_{IR}=$ fixed) when 
the chemical potential is taken as $\mu_{_N} \to 0$ in a specific manner.
From \eqn{sol8} and \eqn{pol3} it is 
\bea
s\sim  T^{d-\theta_{lif}\over a_{lif}}, 
~~~~T\sim r_h^{a_{lif}}\sim0, ~~~
\mu_{_N}\sim  r_h^{2p-14\over p-5}\sim 0,
~~~~\rho= {\rm fixed}
\eea
Especially for $p=3$ case we have
\bea
s\sim  T^{1\over 3}, 
~~~~T\sim r_h^3, ~~~
\mu_{_N}\sim  r_h^4,
~~~~\rho= {\rm fixed}.
\eea
which matches with the result \cite{hs10}. 
Since horizon size vanishes  in this limit
this is an extremal limit.

It is useful to note from the table \eqn{Table2} that 
the dynamical exponents of time, $a_{lif}$, for these Lifshitz 
geometries are all positive definite and generally $a_{lif}>a_{rel}$. 
But also a very interesting observation follows. For a given Lifshitz
D$p$-brane type (electric or magnetic)  
the exponent of $T$ in entropy expression \eqn{sol8} is universally
fixed by the unique fraction 
 ${1 \over \tilde p}$, see the table \eqn{Table2}, where  $\tilde p$ is 
 the number of spatial directions of the corresponding
{\em electric/magnetic} dual D$\tilde p$-brane.
This distinct Hodge-dual behavior of the thermal
entropy of the Lifshitz system
at low temperatures is remarkably present for all the D$p$ solutions. 
Therefore the entropy of the thermal 
 Lifshitz system given in \eqn{sol8}
can also be written as a simple expression
\be\label{sol9}
s_{(p)}\sim T^{1\over \tilde p}.
\ee
Thus for example if $p=1$, we would take $\tilde p=5$,   
for $p=2$, we should take $\tilde p=4$,   
for $p=3$ (self-dual), we should take $\tilde p=3$, 
and for $p=4$, we should take $\tilde p=2$,   
and so on. We get 
the empirical identity
\be
{\tilde p} =6-p={a_{lif}\over d-\theta_{lif}}
\ee
which is indeed true.

The same behaviour as \eqn{sol9} is also  seen in the case of 
M-theory Lifshitz type solutions in the next section. 

\subsection{Lifshitz solutions in eleven dimensions}

There do exist Lifshitz solution in M-theory as well, 
obtainable from `vanishing horizon double 
limits' of corresponding `boosted black M2-branes' \cite{hs10},
\bea\label{sol432}
ds^2_{lif_{M2}}
&=& { r^2}\left( -dx^{+}dx^{-}+{\beta^2\over 4 r^3}
(dx^{-})^2
+dy^2\right) +{1\over 4}{d r^2\over   r^2}  + d\Omega_7^2 ,
\eea
with 4-form field strength $F_4$ being an `electric type flux'. We 
should call them electrically charged Lifshitz membrane solutions.
These solutions have dynamical exponent of time as $a_{lif}={5\over2}$. 
Note that $x^-$ should be taken to be compact 
and one could take it to be the 11-$th$ circle
of M-theory. 
The boundary theory would be  a $1+1$ dimensional CFT. 
The value of $\theta$ can be determined by going to 
the Einstein frame in noncompact directions spanned 
by the coordinates $(x^+,y,r)$, 
and it is given below in the table. 

Similarly  double limits of boosted black  M5-branes  give us
following `magnetically charged' Lifshitz M5 solution 
\bea\label{sol435}
ds^2_{lif_{M5}}
&=& { r^2}\left( -dx^{+}dx^{-}+{\beta^2\over4  r^6}
(dx^{-})^2
+dy_1^2+\cdots+dy_4^2\right) +{4}{d r^2\over  r^2}  + d\Omega_4^2 \nnu\\
\eea
where the $F_4$ flux is taken along $S^4$. These M5-brane Lifshitz
vacua have dynamical exponent $a_{lif}=4$ and the  boundary
theory is $(1+4)$-dimensional CFT. 

Of course, the two Lifshitz vacua \eqn{sol432} and \eqn{sol435}
in M-theory ought to be rightfully seen as
 electric-magnetic (Hodge) dual of each other. 
These  $1/4$ BPS (extremal)
solutions  describe  boundary theories at respective  Lifshitz fixed points.
Making these solutions slightly off-extremal, 
that is including black holes in 
the IR region of the solutions,
we could study the behaviour of their thermal CFTs. The respective 
thermal entropies are summarised as;
\begin{table}[h]
\begin{center} 
\begin{tabular}{ccccc}
\hline  M$p$-brane &$d$ & $a_{lif}$& $ \theta_{lif}$ & 
$s\sim T^\alpha $ \\ \hline 
 \hline
M2&1 & ${5\over 2}$&  ${1\over 2} $&$T^{1\over5}$ \\ \hline
M5&4 & 4 & 2 & $T^{1\over2}$ \\ \hline
\end{tabular} 
\caption{\label{Table3} Dynamical scaling exponents of the 
Lifshitz M2 and M5 
solutions}
\end{center}
\end{table}

As discussed above that M2 and M5 Lifshitz vacua are electric-magnetic dual 
of each other in the same sense as ordinary relativistic  
M2-brane is Hodge-dual to M5-brane and vice versa.
As a curious observation we find that for M-theory Lifshitz solutions
the expressions of the entropy
are  
\bea
&& s_{lif_{M2}} \sim T^{1\over 5}  ~~~~~~~{\rm for~~~ M2} \br
&& s_{lif_{M5}} \sim T^{1\over 2}  ~~~~~~~{\rm for~~~M5} 
\eea
If we pair them up, the expressions
 could then be summarised simply by an expression 
\be
s_{lif_{M(p)}}\sim T^{1\over \tilde p}
\ee
where $\tilde p$ is to be taken as the integer number 
counting the spatial world-volume directions of   dual
M$\tilde p$-brane. 
For example, for M2-brane, $p=2,~\tilde p=5$ and vice versa.
Not only this, we can see this effect of Hodge-duality 
in the case of D$p$-brane Lifshitz vacua too. 
   
On the other hand
 at the (relativistic) UV fixed point, the thermal entropy of the CFTs
goes as
\be
s_{(p)}\sim T^{\alpha_p} 
\ee where $ \alpha_p \ge 2$ and  
 is usually a growing number as $p$ increases, generally
 for all $p$-branes in ten or eleven dimensions. 
 
\section{Interpolating Solutions}
In this section we first take up the issue of bad UV 
behaviour of our Lifshitz geometries of the last sections. 
Then we propose a remedy
so as to regularise these solutions in order to include 
proper AdS metric in the UV region.

\subsection{Problem with Lifshitz solutions in the UV region}

As we  noted, once lightcone coordinate $x^{-}$ is compactified, i.e.
 $x^{-}\sim x^{-} + 2\pi r^{-}$, the Lifshitz 
geometries \eqn{sol3}  do
provide a valid holographic description of a $p$-dimensional nonrelativistic
  CFT, but only in a finite $z$ (energy) range. 
These solutions cannot be trusted in the far UV region.
For example, let us take the D3 case, the string metric 
in this case cannot be trusted near the  boundary 
(UV region) because the physical size of  $x^{-}$ circle 
\be
{R^{-}_{phys}\over l_s}={R_3\over l_s} \beta{r^{-}z} 
\ee
 becomes sub-stringy when $z\to 0$. This is true for all other Lifshitz
like solutions given in \eqn{sol3}, \eqn{sol432} and \eqn{sol435}. 
Thus this UV problem 
exists whenever $x^-$ is a circle! 
There are a few possible ways to tackle this  problem however.
\ben 
\item Of course, standard thing we could do is to include higher 
derivative  (world-sheet) corrections 
to the IR Lifshitz solutions  when the size of $x^-$
 starts becoming sub-stringy.  

\item
Alternatively, as  suggested in Ref. \cite{malda} for the Schrodinger type
solutions, 
 it will be appropriate to go over to 
a T-dual type II string  picture where the T-dualised $x^-$ circle 
will have a finite size.  

\item The third possibility
  could be that, it is quite plausible, to regularize the Lifshitz solutions 
so as to include appropriate boundary (UV) configuration, such 
as discussed  in \cite{hsuv} for the D3 case. 
\een
In general, we naively expect a boundary  Lifshitz theory 
to flow towards becoming a relativistic  theory at  high energies. 
Hence, we can think of attaching  suitable 
boundary configuration, like the conformally AdS geometry such as
in eq.\eqn{sol2d03},  
to the Lifshitz solutions \eqn{sol3}. This we can do for  all $p$ cases. 
\footnote{This can  be achieved by making a  shift
$x^{+}\to x^{+}- x^{-}$ in the above Lifshitz solutions.}
Such D3 brane solutions with regularized UV behaviour become \cite{hsuv}
\bea\label{sol2d01g} 
&&ds^2_{D3}=R_3^2\bigg[ ({1\over z^2}+ {\beta^2 z^{2}}) (dx^{-})^2 
+{-dx^{+}dx^{-} +d\vec x_{(2)}^2\over z^2} +{dz^2\over z^2} + 
d\Omega_{(5)}^2 \bigg] \br && e^\phi=(2\pi)^{-1}g_{YM}^2 \ , ~~~ 
F_{(5)}=4 R_3^4 (1+\star)\omega_{5} \eea 
It should be noted that the solutions \eqn{sol2d01g}
 no longer have the  
asymmetric scaling properties possessed by the purely 
Lifshitz solutions \eqn{sol3}. 
These interpolating solutions \eqn{sol2d01g} behave like 
 solitons which interpolate 
between the IR Lifshitz and the UV relativistic fixed points. 
Namely, in the deep IR region $(z\sim\infty)$ 
it flows  towards a Lifshitz fixed point described by
 $a=3 , \theta=1$. 
The thermal entropy of the $2+1$ boundary CFT
at  IR fixed point  behaves as
\be
 s \sim v_2 (2\pi r^-) T^{1\over 3}\ , \ee
see the table \eqn{Table2}.    
This is an entirely expected behaviour. For example, this behaviour
automatically emerges when vanishing horizon
 double limits are employed on the thermal quantities in thermal CFT 
\cite{hs10, hs10a}. 
While in the deep UV region, as $z\sim 0$, the solution \eqn{sol2d01g}
tends to become a conformally AdS configuration with
 $a=1,~\theta=-1$. 
The thermal entropy of the $2+1$-dimensional CFT
at the UV fixed point behaves as
\bea
S\sim v_2 (2\pi r^-) T^3
\eea    
which is an expected behaviour of a relativistic 3D CFT, 
see the table \eqn{Table1}. 
(Note that we have $r^-$ in the above expressions because
 the coordinate $x^{-}$ is compact having  radius $r^-$.).  
Thus we have an interpolating soliton solution 
of type II string theory which takes us from a Lifshitz solution in IR to 
a relativistic 
solution in UV. That is, the  Lifshitz theory at the  IR
fixed point also has a needed UV completion in terms of relativistic 
 fixed point. This appears to be  true at least
in the supersymmetric examples considered here,
 although it may not be entirely true 
when there is no supersymmetry in the system. 
\footnote{ The  flows from  Lifshitz  solutions 
have been studied earlier  in  suitable phenomenological settings
by \cite{ Bertoldi:2010ca,Braviner:2011kz, Liu:2012wf}. 
We thank the anonymous referee for the information.}

 \subsection{ Interpolating D$p$ solutions }
It is worth while to write down the interpolating  
solutions for all $p$-branes, which behave 
like a Lifshitz solution eq.\eqn{sol3} in the extreme IR and as a relativistic
solution eq.\eqn{sol2de1} in the far UV region. 
The interpolating soliton solutions 
can be written as (for $p\ne 5$)
\bea\label{int03}
ds^2_{Int}
&=&R_p^2 z^{p-3\over p-5} \bigg[ 
\{({1\over z^2}+ {\beta^2\over z^{4/(p-5)}}) (dx^{-})^2 +{-dx^{+}dx^{-}
+d\vec x_{(p-1)}^2\over z^2}  
+{4\over (5-p)^2} {dz^2\over  z^2}\}  + d\Omega_{(8-p)}^2 \bigg] \br
&=&R_p^2 z^{p-3\over p-5} \bigg[ 
\{ {K\over z^2} (dx^{-})^2 +{-dx^{+}dx^{-}
+d\vec x_{(p-1)}^2\over z^2}  
+{4\over (5-p)^2} {dz^2\over  z^2}\}  + d\Omega_{(8-p)}^2 \bigg] \br
e^\phi&=&(2\pi)^{2-p}g_{YM}^2R_p^{3-p}{ z^{(7-p)(p-3)\over 2p-10}}
\eea 
with the $(p+2)$-form flux. 
Where the new function 
\be\label{int032}
K(z)=1+{1\over 4} \left({z\over z_{IR}}\right)^{2p-14\over p-5}
\ee 
is also an harmonic function and plays the role of the interpolating function.
The parameter $z_{IR}>0$ is
an intermediate IR scale and can be related to $\beta$. It is 
being called  interpolating solution because
the metric \eqn{int03} 
smoothly connects Lifshitz and AdS regions, even when $x^-$ is compact. 
It is much like a `wormhole' geometry, the size of $x^-$ circle stays finite. 
In the asymptotic UV region $(z\ll z_{IR})$ where $K\approx 1$,
it  starts behaving relativistically, 
while for $z\gg z_{IR}$  where 
$K\approx ({z\over z_{IR}})^{2(p-7)\over p-5}$ it 
behaves like a Lifshitz spacetime. Note that,
since these solutions are interpolating solitonic configurations 
any scaling symmetry of the metric 
\eqn{int03} is explicitly broken.
The  {\it  scaling} or dilatation symmetry of
the metric  becomes explicit
 in extreme IR or  UV regions only. 
This interpolating geometry is depicted schematically in
 the figure \eqn{figure3}.
\begin{figure}[h]
\centerline{\epsfxsize=5in
\epsffile{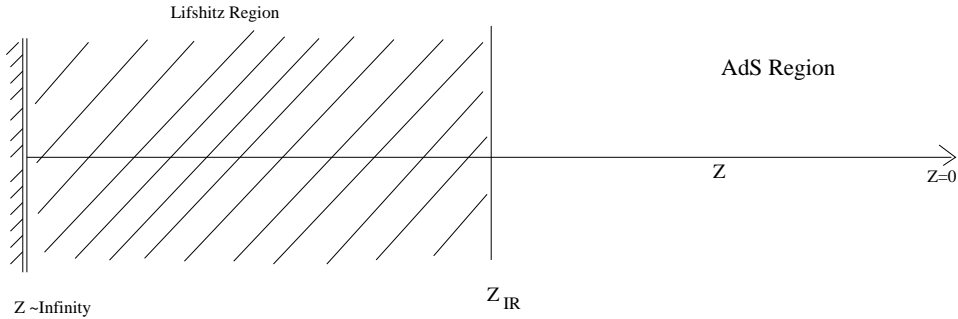} }
\caption{\label{figure3} In zero temperature solutions the Lifshitz  
window (the shaded region) starts at $z\sim \infty$ $(r\sim 0)$ and ends at
$z_{IR}$.}
\end{figure}

The explicit
compactification of the metric \eqn{int03} gives a $(p+1)$ dimensional
spacetime
\bea\label{int035}
&&ds^2_{p+1}
=L^2 z^{2(p^2-7p+14)\over (p-1) (p-5)}K^{1\over p-1} \bigg[ 
 -{(dx^{+})^2\over 4 z^2 K}+ {d\vec x_{(p-1)}^2\over z^2}  
+{4\over (5-p)^2} {dz^2\over  z^2}   \bigg]   
\eea 
where $K$ is  given above in \eqn{int032}.
There is  a running $(p+1)$-dimensional dilaton field
\bea
&& e^{-2\phi_{(p+1)}}\sim z^{p-5\over2} \sqrt{K} \br
&&A_{(1)}= -{1\over 2K}dx^{+}
\eea
where $L^2$ is an  specific size factor which follows from compactification.

It is also plausible to include  black holes in these 
interpolating solutions \eqn{int03}. 
This can be done systematically by employing the boost, 
see \cite{hs10a},
 and only changes occur in the spacetime metric
\bea\label{int033}
ds^2_{Lif}&=&R_p^2 z^{p-3\over p-5} \bigg[ 
\{-{f(dx^{+})^2\over4 z^2 K}+ 
{d\vec x_{(p-1)}^2\over  z^2}  
+{4\over (5-p)^2} {dz^2\over f z^2}\}
+{K\over z^2}(dx^- - A)^2
  + d\Omega_{(8-p)}^2 \bigg]\nonumber \\
\eea
where 1-form $$A\equiv {(1+f)+\lambda^{-2}(1-f)\over 4K}dx^{+}$$ 
and the harmonic functions 
 \bea\label{int034}
f(z)&=&1-\left({z\over z_{0}}\right)^{2p-14\over p-5}\br
K(z)
&=& 1+{\lambda^2-1\over 4\lambda}\left({z\over z_{0}}\right)^{2p-14\over p-5} 
\equiv 1+{1\over 4}\left({z\over z_{IR}}\right)^{2p-14\over p-5}
\eea
The dilaton and other form fields remain unchanged.
The $z=z_0$ is the location of the black hole horizon.
Note that $\lambda$ is the boost parameter in the above.
In the absence of boost, $\lambda=1$, then $K=1$.
Since the Lifshitz region for many physical applications would 
be some intermediate (IR) region, it would be worth while 
to take $z_0 > z_{IR}>0$, and this is always guaranteed from \eqn{int034}. 
In this way, the black hole singularity is capped by its  horizon.
 We  call the intermediate region $z_0\ge z\ge z_{IR}$
as the  Lifshitz window region
where parameter $z_{IR}$ provides the effective width of the 
window beyond the horizon. 
 While in the deep UV region, $z\ll z_{IR}$ the solutions 
become asymptotically conformally AdS, see the figure \eqn{figure1}.
 \begin{figure}[h]
\centerline{\epsfxsize=5in
\epsffile{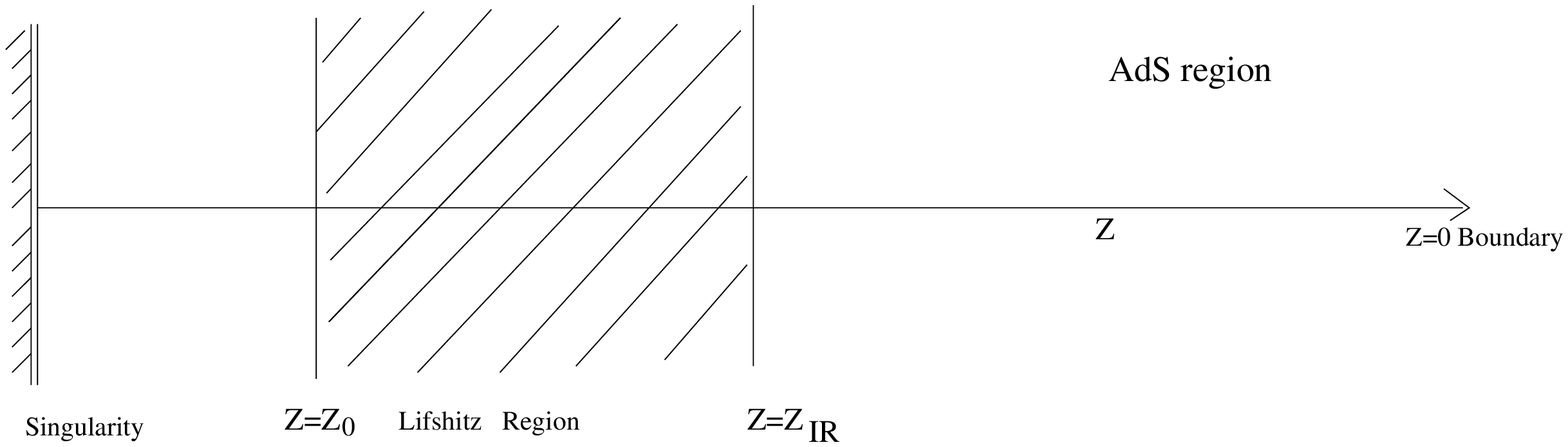} }
\caption{\label{figure1} The Lifshitz  
window appears as the shaded region. It starts at $z_0$ and ends at
$z_{IR}$.}
\end{figure}
Note that the size of Lifshitz window  depends on the boost, it
can be widened if we take 
$\lambda$  sufficiently large. Specially if $\lambda=1$ 
the Lifshitz region altogether disappears and we get ordinary 
AdS black hole solutions.
The  Lifshitz BH solutions \eqn{int033} with an 
intermediate Lifshitz region should
present a good IR description (at finite temperature)
of a boundary Lifshitz theory. 
The black hole horizon provides an effective
 IR (thermal) cut-off scale in the dual CFT.

\subsection{Entanglement Entropy}

In order to find the entanglement entropy of the CFT, we shall use the 
interpolating zero temperature solutions like \eqn{int03} or \eqn{int035}. 
According to Ryu-Takayanagi proposal \cite{RT}, if we pick up a
subsystem A  (with its boundary 
$\partial A$), the subsystem has an entanglement of its states
with its parent system.
Then the entanglement entropy of the subsystem A 
 can be given geometrically 
in terms of the area of an extremal surface $X_{(p-1)}$ 
(space like $(p-1)$-dimensional surface) ending on to 
the boundary $\partial A$. Thus we have
\be S_{Ent}(A)={1\over 4G_{p+1} }[Area]_X
\ee 
The extremal surface $X$ extends well inside the bulk geometry. 
We pick up the 
subsystem $A$ to be a rectangular strip
 along $x_1(z), x^-(y)$ at any fixed time. 
Note that, $x^+$ is identified with boundary
time coordinate and it does not depend upon $y$. Also 
as per our study we have  to take $x^-$ being  a compact coordinate. 
The range of the coordinates is 
$-l/2 \le x_1\le l/2$ and the regulated size of
 other coordinates is $0\le x^i\le l^i$.
(For noncompact $x^-$ the subsystem $A$  must be thought off as a strip  
of finite width $l$ stretched along  spatial direction $x^-$.) 
For our calculations we shall
 consider the $(p+1)$-dimensional Einstein metric as in \eqn{int035}. 
Then 
\be
S_{Ent}={1\over 4 G_{p+1}}\int \sqrt{g_X}
\ee
where $g_X$ is the induced metric on the $(p-1)$-dimensional
extremal surface $X$. Note after the compactification along $x^-$
the strip becomes just an interval along $x_1$.
Using the compactified metric \eqn{int035}, we find that 
\be\label{kl1}
S_{Ent}={V_{p-2}L^d\over 2 G_{p+1}}
\int_{z_\ast}^{z_\infty}dz ~ z^{9-p\over p-5} 
\sqrt{K}\sqrt{{4\over (5-p)^2} + (x_1')^2}
\ee  
where $z_\infty\approx 0$ is the UV cut-off and $z_\ast$ is the turning point.
$V_{p-2}$ is the size of the ensemble box stretched along rest of the spatial
directions, $x_2,\cdots, x_{p-1}$.   $K$ is as given in \eqn{int032}.  
The extremal surface satisfies the first order equation
\be\label{kl3}
{dx_1\over dz}= {2\over 5-p} {C z^{p-9\over p-5}\over
\sqrt{1+{1\over 4}({z\over z_{IR}})^{2(p-7)\over p-5}
-C^2 z^{2(p-9)\over p-5}}}
\ee 
where $C$ is the integration constant. The turning point arises 
 where $x_1'|_{z_\ast}=\infty$. While near the boundary point 
$x_1'|_{z_\infty}\sim 0$. 
Finding solutions of first order
differential equation \eqn{kl3} is much like solving a classical orbit in
 the central force problem with given boundary (initial) conditions. 
The term  $C^2 z^{2(p-9)\over p-5}$ plays the role of a repulsive 
centrifugal type force, while the term
$-{1\over 4}({z\over z_{IR}})^{2(p-7)\over p-5}$ behaves like an attractive 
central force. Thus Lifshitz deformation in the IR region  is
 of attractive nature while the repulsive forces mainly 
come from the curvature of AdS spacetime. 
This gives  finally the entropy formula
\be\label{fg3}
S_{Ent}={V_{p-2}L^d\over 2 G_{p+1}}
\int_{z_\ast}^{z_\infty}dz ~z^{9-p\over p-5} {2\over (5-p)} 
{1+{1\over 4}({z\over z_{IR}})^{2(p-7)\over p-5}
\over\sqrt{ 1+{1\over 4}({z\over z_{IR}})^{2(p-7)\over p-5}
-C^2z^{2(p-9)\over p-5}}}
\ee  
This expression matches with other calculations in the literature 
\cite{narayan12, narayan13}. If we set $1/z_{IR}$ to be zero, the expression 
\eqn{fg3} 
reduces to the entanglement entropy in the relativistic CFT system.
It can be seen that the turning point of the extremal surface
in the purely AdS case appears at the value
$z=z_c\equiv C^{5-p\over p-9}$. Thus we always have $z_\ast >z_c$ 
for the Lifshitz system. Thus the area of the entremal surface is larger
in the Lifshitz case. 
Hence the entanglement entropy of the Lifshitz system
is generally larger compared to the relativistic (AdS) case.  That is
\be
S_{Ent}^{Lifshitz} > 
S_{Ent}^{AdS}  .
\ee

At the finite temperature, looking at eqs. \eqn{int033} and \eqn{int034}, 
we find that
\be\label{kl1a}
S_{Ent}={V_{p-2}L^d\over 2 G_{p+1}}
\int_{z_\ast}^{z_\infty}dz ~ z^{9-p\over p-5} 
\sqrt{K}\sqrt{{4\over (5-p)^2 f} + (x_1')^2}
\ee  
where $f(z)$ is given earlier in \eqn{int034}. We always have $z_0 > z_{IR}$.
The extremal surface satisfies the first order equation
\be\label{kl3a}
{dx_1\over dz}= {2\over 5-p}{1\over\sqrt{f}} {C z^{p-9\over p-5}\over
\sqrt{1+{1\over 4}({z\over z_{IR}})^{2(p-7)\over p-5}
-C^2 z^{2(p-9)\over p-5}}}
\ee 
This gives  finally the entanglement entropy formula (at finite temperature)
\be\label{fg3a}
S_{Ent}={V_{p-2}L^d\over 2 G_{p+1}}
\int_{z_\ast}^{z_\infty}dz ~z^{9-p\over p-5} {2\over (5-p)}{1\over\sqrt{f}} 
{1+{1\over 4}({z\over z_{IR}})^{2(p-7)\over p-5}
\over\sqrt{ 1+{1\over 4}({z\over z_{IR}})^{2(p-7)\over p-5}
-C^2z^{2(p-9)\over p-5}}}\ .
\ee

\section{Conclusion}
We have presented  quarter BPS Lifshitz 
D$p$-brane vacua and obtained explicit expressions for
 their thermal quantities at finite temperature. We studied how
 various
quantities behave if the low temperature limit is taken, at fixed
 charge density. We also studied 
how Lifshitz D$p$-brane systems are mapped under
 electric-magnetic duality. For example 
the entropy of the near extremal Lifshitz D$p$-brane goes as
\be S_{(p)}\sim T^{1\over \tilde p}
\ee
where $\tilde p$ is an integer giving us 
the number of spatial world-volume 
directions of the magnetic dual Lifshitz D$\tilde p$-brane. Thus
\be
{\tilde p} =6-p={a_{lif}\over d-\theta_{lif}}.
\ee
Surprisingly, the same behaviour persists also  for the  extremal 
Lifshitz M2 and M5-brane vacua, which are electric-magnetic 
duals of each other in M-theory. 
Thus the Lifshitz  systems though being inherently nonrelativistic
do encode deep quantum relationships 
such as electric-magnetic duality. Any measurement of these Lifshitz
thermal exponents, say $
s\sim T^{1\over4},~
s\sim T^{1\over3}$ or
$s\sim T^{1\over2}$
 in condensed matter systems with 1, 2 or 3 spatial 
dimensions, respectively, could be taken as a signature test 
of electric-magnetic (Hodge) duality in nonrelativistic string systems.
It would also be useful to further understand the basic reason behind it. 

We have written down the interpolating solutions as well. These class of
 solutions are well behaved and can be trusted for the classical analysis
in the UV region also.
 The entanglement entropy  is calculated by using these interpolating
solutions and its expression matches with the 
 recent works of \cite{narayan13, narayan12}.
We also find that the entanglement entropy of the Lifshitz system
is generally larger compared to the relativistic (AdS) case.

\vskip.5cm
\noindent{\it Acknowledgments:}
I wish to thank Shibaji Roy for useful discussions.

\vskip1cm
\centerline{----------------------------}

\end{document}